\documentclass{article}
\usepackage{amsmath,amsfonts,latexsym,graphicx,amssymb}
\def\dST{\displaystyle}

\newtheorem{lemma}{Lemma}
\def\1{{\mathchoice{\rm 1\mskip-4mu l}{\rm 1\mskip-4mu l}%
{\rm 1\mskip-4.5mu l}{\rm 1\mskip-5mu l}}}
\def\<{\langle}
\def\>{\rangle}

\def\dST{\displaystyle}
\def\ket#1{|\kern0.7pt #1\>}
\def\bra#1{\langle #1\kern0.7pt|}
\def\scalar#1#2{\langle#1\kern0.7pt|\kern0.7pt #2\rangle}
\def\Tr{{\rm Tr}\kern1pt}

\title{Discord Derived from Tsallis Entropy}
\author{Jacek Jurkowski\\
Institute of Physics, Nicolaus Copernicus University\\
 ul. Grudzi\c{a}dzka 5/7, 87--100 Toru\'n, Poland}
\date{}

\begin{document}

\maketitle

\begin{abstract}
Due to some ambiguity in defining mutual Tsallis entropy in the classical probability theory, its generalization
to quantum theory is discussed and, as a consequence,
two types of generalized quantum discord, called $q$-discords, are defined in terms of quantum Tsallis entropy.
$q$-discords for two-qubit Werner and isotropic states are calculated and it is shown that one of them is positive, at least for
states under investigation, for all $q>0$. Finally, an analytical expression for $q$-discord of certain
family of two-qubit X states is presented. 

\end{abstract}

\section{Introduction}

There is a still growing interest in quantifying and measuring correlations in composite quantum systems (for
a recent review see \cite{IJQI}). Such notions as entanglement, quantum discord and other entropic measures of quantumness
proved to be useful in these investigations \cite{QIT,Hor09,Zurek,Vedral,HHH,Ros}. While entanglement from a very definition reflects by itself some
specific structure of the density matrix representing the state, discord measures the amount of quantum correlations  between subsystems in terms of mutual von Neumann entropy. 

On the other hand, for many years the nonadditive generalizations of Shannon and von Neumann entropy functions, such as Renyi   entropy \cite{Renyi}, Tsallis entropy \cite{T-1} and others \cite{Ros,Masi},
has been used to describe some aspects of statistical and thermodynamic behaviour of systems far from equilibrium.
These families of entropy functions parametrized by $q>0$ seem to incorporate in a natural manner some correlations 
present in composite systems. In what follows we focus on Tsallis entropy which, except additivity, displays 
similar properties (nonnegativity, concavity) to those of von Neumann entropy \cite{Furu,Furu2,Hu} and therefore, in most applications in the field of quantum information, can be used as well 
\cite{Abe0,Abe,Yam,GuhLew,Kim}. In particular, an observation due to Abe \cite{Abe} that using Tsallis entropy
one can detect more separable two-qubit Werner states than using Shannon entropy  seems very promising.

In this article, we investigate the generalization of mutual entropy and quantum discord when defined in terms
of Tsallis entropy. It turns out that even at the classical level this generalization can be done following two paths. Hence, at the quantum level  
we introduce two quantities, called $q$-discords $D_q(A:B)$ and $\widetilde{D}_q(A:B)$, which generalize quantum discord, and we investigate them for Werner and isotropic states. We give some arguments that the use of $q$-mutual entropy derived from 
$q$-expectation value results in 
the $q$-discord function which, at least for Werner and isotropic states, reflects in a more appropriate way
the amount of quantum correlations for all values of $q>0$ than the other one. 

Finally, we give an example indicating that both types of $q$-discord can take negative values for some specific quantum states belonging to the family of two-qubit X states \cite{Ali}.

\section{Classical Information Theory}

Let us start we some basics of classical information theory, where, in fact, 
some ambiguities crucial for further analysis arise. 

\subsection{Conditional and mutual Shannon entropy}

Let us consider two classical  systems
$A$ and $B$ described by discrete probability spaces 
$(\Omega_A=\{A_1,A_2,\ldots\},P_A=(p_i^{(A)}))$ and 
$(\Omega_B=\{B_1,B_2,\ldots\},P_B=(p_j^{(B)}))$, respectively. 
Information contained in any probability distribution $P=(p_1,p_2,\ldots)$ 
can be quantified by the Shannon entropy function
\[
S(P)=-\sum_ip_i\log p_i.
\]
Consider now a joint probability space $(\Omega=\Omega_A\times
\Omega_B,P)$, where $P=\{p_{ij}\}$ is the joint probability, and define 
conditional probability $p_{i|j}$  as
\begin{equation}\label{cond-prob}
p_{i|j}:=\frac{p_{ij}}{p^{(B)}_j}\,.
\end{equation}
Hence, Shannon entropy of 
conditional probability (\ref{cond-prob}) reads
\[
S(A|B_j):=-\sum_ip_{i|j}\log p_{i|j}.
\]
Averaging it over subsystem B leads to the notion of \textit{conditional entropy}  
\begin{equation}\label{S-cond-e}
S(A|B):=\mathbb{E}^{B}[S(A|B_j)]=\sum_jp^{(B)}_jS(A|B_j),
\end{equation}
which is an expectation value of entropy of conditional 
probability with respect to marginal probability $P_B$.
As an obvious consequence,
\begin{equation}\label{S(A|B)}
S(A|B)=S(AB)-S(B),
\end{equation}
with $S(AB):=-\sum_{ij}p_{ij}\log p_{ij}$  being the joint entropy, holds.

The amount of information on the subsystem A contained in B is characterized by 
\textit{mutual entropy}
\begin{equation}\label{mutual-S-entropy}
I(A:B):=S(A)-S(A|B).
\end{equation}
Due to (\ref{S(A|B)}) it follows immediately that 
$I(A:B)$ can be alternatively described by
\begin{equation}\label{mutual-S-entropy-2}
I(A:B)=S(A)+S(B)-S(AB),
\end{equation}
as long as the Shannon entropy function is used. As we will argue in Sect.~\ref{CcMTe} 
this equivalence is no longer valid when using more general entropy function, i.e., Tsallis entropy.

\subsection{Classical conditional and mutual Tsallis entropy}
\label{CcMTe}

Using generalized logarithmic function, called \textit{$q$-logarithm},
\[
\ln_q x\;=\;\frac{x^{1-q}-1}{1-q}
\]
one can define \textit{Tsallis entropy} of rank $q$ as
\[
T_q(P)\;:=\;-\sum_jp_j^q\ln_qp_j\;=\;\frac{1}{q-1}\Big(1-\sum_jp_j^q\Big)
\]
and related to it, \textit{Tsallis entropy of conditional probability}
\begin{equation}\label{cond-entropy}
T_q(A|B_j)\;:=\;-\sum_ip_{i|j}^q\ln_qp_{i|j}\;=\;\frac{1}{q-1}\Big(1-\sum_ip_{i|j}^q\Big)\,.
\end{equation}
Note that $q\to1$ corresponds to Shannon entropy.
Now, in order to define \textit{Tsallis conditional entropy} one can follow several nonequivalent  ways.
\begin{itemize}
\item[1)] In a perfect analogy to (\ref{S-cond-e}) one could define
\[ {\cal T}_q(A|B):=\mathbb{E}^{B}[T_q(A|B_j)]=\sum_jp^{(B)}_jT_q(A|B_j),
\] 
where $\mathbb{E}^B$ means the expectation value with respect to the marginal 
probability $P(B)$.
Unfortunately, this quantity has no useful properties, in particular, there is no relation similar to (\ref{S(A|B)})!  
\item[2)] If  someone wants the property (\ref{S(A|B)}) to hold then an expectation value with
modified weights $(p^{(B)}_j)^q$ should be considered, i.e.,
\begin{equation}\label{cond-zla}
\widetilde{T}_q(A|B):=\<T_q(A|B_j)\>^B_q:=\sum_j(p^{(B)})^q_jT_q(A|B_j)=-\sum_{i,j}p_{ij}^q\ln_q\frac{p_{ij}}
{p^{(B)}_j}\,.
\end{equation}
As a consequence of the following property of $q$-logarythm:
\begin{equation}\label{q-ln}
\ln_q\frac{x}{y}=\ln_qx-\Big(\frac{y}{x}\Big)^{q-1}\ln_qy
\end{equation}
one easily obtains that
\begin{equation}\label{T(A|B)}
\widetilde{T}_q(A|B)=\widetilde{T}_q(AB)-\widetilde{T}_q(B)\,,
\end{equation}
but $(P^{(B)})^q$ is not a probability distribution, hence $\<\,\cdot\,\>_q$ is not a legitimate expectation value.  
\item[3)] To avoid this,  one defines 
a $q$-\textit{expectation value} of any random variable $f=(f_1,f_2,\ldots)$ and 
probability $P=(p_1,p_2,\ldots)$ as
\[
\mathbb{E}_q[f]\;:=\;\frac{\sum_jp_j^qf_j}{\sum_jp_j^q},
\]
which obviously preserves unity, $\mathbb{E}_q[1]=1$. Usually, the probability distribution 
\[ r_i=\frac{(p_i)^q}{\sum_j(p_j)^q}
\]
is refered to as an \textit{escort probability}. Hence, conditional Tsallis entropy $T_q(A|B)$  can be defined as the $q$-expectation value with respect to the marginal
probability $P(B)$ as
\begin{equation}\label{cond-entropy-2}
T_q(A|B):=\mathbb{E}^{B}_q[T(A|B_j)]=\sum_jr^{(B)}_jT_q(A|B_j)=\frac{\sum_i(p_i^{(B)})^qT_q(A|B_i)}
{\sum_j(p^{(B)}_j)^q}
\end{equation}
in perfect  analogy to the case of Shannon entropy (\ref{S-cond-e}), where
\[ S(A|B)\;=\;\mathbb{E}^B[S(A|B_j)]\,.
\]
Now,  (\ref{T(A|B)}) is no longer valid, but we can rewrite (\ref{cond-entropy-2}) using (\ref{cond-entropy}) and the relation $p_{ij}=p^{(B)}_jp_{i|j}$  as
\begin{eqnarray*}
T_q(A|B) &=& -\frac1N\sum_j(p^B_j)^q\sum_i(p_{i|j})^q\ln_q p_{i|j} \\
&=& -\frac1N\sum_{i,j}p_{ij}^q\ln_q\frac{p_{ij}}{p^B_j}\,,
\end{eqnarray*}
where
\[
N\;=\;\sum_j(p_j^B)^q\;=\;1+(1-q)T_q(B)\,.
\]
Using (\ref{q-ln}) results in
\begin{eqnarray*}
T_q(A|B) &=&  -\frac1N\sum_{i,j}p_{ij}^q\Big[\ln_qp_{ij}-\Big(\frac{p^B_j}{p_{ij}}\Big)^{q-1}
\ln_qp^B_j\Big] \\
&=& -\frac1N\sum_{i,j}\Big[p_{ij}^q\ln_qp_{ij}-p_{ij}(p^B_j)^{q-1}\ln_qp^B_j\Big] \\
&=& \frac{T_q(A,B)-T_q(B)}{N}\,.
\end{eqnarray*}
Finally, one arrives at the modified relation (\ref{T(A|B)}), i.e.,
\begin{equation}\label{T_q-cond}
T_q(A|B)=\frac{T_q(AB)-T_q(B)}{\sum_j(p^{(B)}_j)^q}=\frac{T_q(AB)-T_q(B)}{1+(1-q)T_q(B)}\,.
\end{equation}
\end{itemize}

In principle, both (\ref{T(A|B)}) and (\ref{T_q-cond}) can be used to define \textit{mutual 
Tsallis entropy}. Using (\ref{T(A|B)}) leads to
\begin{equation}\label{trad-ME}
\widetilde{\cal I}_q(A:B) = \widetilde{T}_q(A)+\widetilde{T}_q(B)-\widetilde{T}_q(AB)
\end{equation}
and mimic the well-known case of Shannon entropy. On the other hand, if
one consequently defines the mutual information as
\begin{equation}\label{mut-inf}
{\cal I}_q(A:B)\;=\;T_q(A)-T_q(A|B)\,,
\end{equation}
and takes into account (\ref{T_q-cond}), then
\begin{equation}\label{mut-inf-2}
{\cal I}_q(A:B)\;=\;\frac{T_q(A)+T_q(B)-T_q(A,B)+(1-q)T_q(A)T_q(B)}{1+(1-q)T_q(B)}\,.
\end{equation}

Both (\ref{trad-ME}) and (\ref{mut-inf-2}) can be generalized and used to quantify some 
correlations in composite quantum systems but we will argue in Sect.~\ref{QT} that the use of 
(\ref{mut-inf-2}) has some advantages.

\section{Quantum Theory}
\label{QT}

It is   a common agreement that the most suitable quantity to describe correlations between subsystems 
A and B is \textit{mutual entropy} (see (\ref{mutual-S-entropy})) usually defined in terms of von Neumann entropy 
\[ H(AB):=H(\rho_{AB})=-{\rm Tr}(\rho_{AB}\log\rho_{AB})
\]
as
\[
{\cal I}(A:B)=H(A)+H(B)-H(AB)\,,
\]
where $H(A)=H(\rho_A)$, $H(B)=H(\rho_B)$, $H(AB)=H(\rho_{AB})$, $\rho_A={\rm Tr}_B(\rho_{AB})$,
$\rho_B={\rm Tr}_A(\rho_{AB})$ being reduced density matrices.
Clearly, ${\cal I}(A:B)$ contains information about all correlations enclosed in the state $\rho_{AB}$
and one can pose the question which of them  are of classical origin and which one are pure quantum 
\cite{Zurek,Vedral,Vedral-2,Maziero}.
One can decompose (in a nonunique way) total correlations ${\cal I}(A:B)$ as
\[
{\cal I}(A:B)\;=\;\underset{\mbox{quantum}}{D(A:B)}\;+
\;\underset{\mbox{classical}}{C(A:B)}\,.
\]
The classical part $C(A:B)$ can be determined by optimizing the local measurement procedure as proposed 
in \cite{Zurek}. The quantity $D(A:B)$ is then called \textit{quantum discord}. In what follows we 
generalize this notion using Tsallis entropy instead of von Neumann ones.

Here we recall some properties of von Neumann entropy crucial for further analysis \cite{QIT}.
The first one is additivity with respect to the tensor product,
\[ H(A\otimes B)=H(A)+H(B)\,.
\]
Next, the so-called subadditivity property (SA) 
\begin{equation}\label{SA}
H(AB)\leq H(A)+H(B)
\end{equation}
ensures that von Neumann mutual entropy is nonnegative
\[
{\cal I}(A:B)\geq 0\,,
\]
and strong subadditivity (SSA) \cite{QIT,Wehrl}
\[
 H(ABC)+H(B)\;\leq\;H(AB)+H(BC)\,,
\]
which is essentially a tri-partite property,
implies nonnegativity of quantum discord $D(A:B)$ \cite{Zurek}.

In the framework of composite \textit{quantum} systems 
\textit{Tsallis entropy} can be defined in perfect analogy to the classical case as
\[
T_q(AB)\;:=\;T_q(\rho_{AB})\;=\;\frac{1}{q-1}\Big(1-\Tr(\rho_{AB}^q)\Big)\,,\quad q>0,\;q\neq 1\,.
\]
Tsallis entropy function $T_{q}(x)$ is nonnegative, concave and, if $\rho_{AB}$ is pure
then $T_{q}(A)=T_{q}(B)$, but $T_{q}(x)$ is no longer additive with respect
to the tensor product \cite{Abe,Yam,Furu,Furu2}, instead we have pseudo-additivity (PA),
\begin{equation}\label{PA}
T_{q}(A\otimes B)=T_{q}(A)+T_{q}(B)+(1-q)T_{q}(A) T_{q}(B).
\end{equation}
This property, in fact, makes Tsallis entropy useful in nonextensive statistical mechanics \cite{T-2}.
From (\ref{PA}) results that SA fails for arbitrary $q>0$. Note, however, that 
for $q>1$ one obtains
\begin{eqnarray*}
T_q(A\otimes B) &\leq& T_q(A)+T_q(B)
\end{eqnarray*}
and moreover \cite{Aud}
\begin{equation}\label{T-sub}
T_q(AB)\;\leq\;T_q(A)+T_q(B),
\end{equation}
hence SA holds.

Let us define two types of
\textit{quantum mutual Tsallis entropy} \cite{Yam,Furu2}, both measuring total correlations between subsystems $A$ 
and $B$, as 
\begin{eqnarray}\label{mut-inf-3a}
\widetilde{\cal I}_q(A:B) &=& T_q(A)+T_q(B)-T_q(AB)\,, \\
{\cal I}_q(A:B) &=& \frac{T_q(A)+T_q(B)-T_q(AB)+(1-q)T_q(A)T_q(B)}{1+(1-q)T_q(B)}\,,
\label{mut-inf-3b}
\end{eqnarray}
Note that all the objects entering both types of mutual entropy given by (\ref{mut-inf-3a}) 
and (\ref{mut-inf-3b}), are already well-defined. Moreover, (\ref{T-sub}) implies that 
$\widetilde{\cal I}_q(A:B)$ is nonnegative for $q>1$. Unfortunately, we are lacking this property for
${\cal I}_q(A:B)$.

In what follows, we will argue that (\ref{mut-inf-3b}) displays some advantages when describing 
quantum correlations in composite systems.

In order to determine classical correlations ${\cal C}_q(A:B)$ based on Tsallis entropy we follow 
\cite{Zurek}. For this, let us consider a perfect measurement on subsystem $B$, defined by a set of 
one-dimensional projectors $\{\Pi_k\}$, 
yielding  post-measurement states 
\begin{equation}\label{rho_k}
\rho_k\;=\;\frac{1}{p_k}(I\otimes\Pi_k)\rho_{AB}(I\otimes\Pi_k)^\dagger\,,\qquad k=1,2,\ldots\,,
\end{equation}
where 
\begin{equation}\label{p_k}
p_k=\Tr(I\otimes\Pi_k)\rho_{AB}.
\end{equation} 
Let us define \textit{quantum conditional Tsallis entropy} 
with respect to this measurement by
\begin{equation}\label{exp-cond}
T_q(\rho_{AB}|\{\Pi_k\})\;:=\;\mathbb{E}_q^{\Pi}[T_q(\rho_k)]\;=\;
\frac{\dST\sum_k p_k^qT_{q}(\rho_k)}{\dST\sum_kp_k^q}\,.
\end{equation}
In fact, $T_q(\rho_{AB}|\{\Pi_k\})$ is fully characterized by an ensemble $\{\rho_k,p_k\}$.
Now, the information gained about A as a result of measurement on B is   
\begin{equation}\label{class-cor}
{\cal C}_q(A:B)\;=\;\sup_{\{\Pi_k\}}\Big(T_q(A)-T_q(\rho_{AB}|\{\Pi_k\})\Big)\,,
\end{equation}
when optimization is taken over all perfect measurements. One assumes that the quantity ${\cal C}_q(A:B)$ encodes classical correlations.

Finally, we define \textit{$q$-discord} by
\begin{eqnarray}
{D}_q(A:B) &=&{\cal I}_q(A:B)-{\cal C}_q(A:B)\,.
\end{eqnarray}
Note, however, that instead of ${\cal I}_q(A:B)$ one could use $\widetilde{\cal I}_q(A:B)$ given by 
(\ref{mut-inf-3a}) and 
instead of (\ref{class-cor}) one could define 
\begin{equation}\label{class-cor-2}
\widetilde{\cal C}_q(A:B)\;=\;\sup_{\{\Pi_k\}}\Big(T_q(A)-\widetilde{T}_q(\rho_{AB}|\{\Pi_k\})\Big)\,,
\end{equation}
where 
\begin{equation}\label{exp-cond-2}
\widetilde{T}_q(\rho_{AB}|\{\Pi_k\})\;:=\;\mathbb{E}^{\Pi}[T_q(\rho_k)]\;=\;
\sum_k p_kT_{q}(\rho_k)
\end{equation}
involves only the ordinary expectation value.
As a consequence, we obtain another type of $q$-discord
\begin{eqnarray}
\widetilde{D}_q(A:B) &=&\widetilde{\cal I}_q(A:B)-\widetilde{\cal C}_q(A:B)\,.
\end{eqnarray}
In \cite{Coles} it is proven that $\widetilde{D}_2(A:B)\geq0$ but some numerical results suggest 
that $\widetilde{D}_q(A:B)$ can also be negative for 
some values of $q>1$. 

In Sect.~\ref{W-q-d} we determine and comment on both types of $q$-discord for Werner and isotropic states.

\section{$q$-Discord for Werner and isotropic states}
\label{W-q-d}

Let us consider $2\otimes 2$ Werner states $\rho_W(\lambda)$, parametrized by $0\leq \lambda\leq 1$, which are invariant with respect to local unitary transformations, 
i.e., $U\otimes U\rho_W(\lambda)\, (U\otimes U)^\dagger=\rho_W(\lambda)$, defined by
\[
\rho_W(\lambda)\;=\;\frac16[(2-\lambda)I_4+(2\lambda-1)\mathbb{F}]\,,
\]
where $\mathbb{F}$ is the flip operator $\mathbb{F}(\ket{\phi}\otimes\ket{\psi})=
\ket{\psi}\otimes\ket{\phi}$ and $I_4$ is $4\times 4$ identity matrix. 
Straightforward algebra yields
\begin{eqnarray}\label{q-ent-sub}
T_q(A) &=& T_q(B)\;=\;\frac{1}{q-1}(1-2^{1-q})\,, \\
T_q(A,B) &=& \frac{1}{q-1}\Big[1-3\Big(\frac{1+\lambda}{6}\Big)^q-
\Big(\frac{1-\lambda}{2}\Big)^q\Big]\,.\label{q-ent-sys}
\end{eqnarray}
Now, using (\ref{rho_k}) and (\ref{p_k}), one obtains $p_1 =p_2 =\frac12$ 
and 
\[
T_q(\rho_1)=T_q(\rho_2)=T_q(\rho_W(\lambda)|\{\Pi_k\})=\frac{1}{q-1}\Big[1-\Big(\frac{2-\lambda}{3}\Big)^q-
\Big(\frac{1+\lambda}{3}\Big)^q\Big],
\]
where, in fact, $T_q(\rho_W(\lambda)|\{\Pi_k\})$ is independent on a particular measurement, as one should 
expect according to local unitary invariance of Werner states.   Hence,
\[
{\cal C}_q(A:B)=\widetilde{\cal C}_q(A:B)=\frac{1}{q-1}\Big[\Big(\frac{2-\lambda}{3}\Big)^q+
\Big(\frac{1+\lambda}{3}\Big)^q-2^{1-q}\Big]
\]
and the discords $D_q(A:B)$ and $\widetilde{D}_q(A:B)$ differ one from another by total correlation terms only.
The later ones obtained from (\ref{mut-inf-3a}) and (\ref{mut-inf-3b}) using (\ref{q-ent-sub}),
(\ref{q-ent-sys}) read
\begin{eqnarray}
\widetilde{\cal I}_q(A:B) &=& \frac{1}{q-1}\Big[1-2^{2-q}+\Big(\frac{1-\lambda}{2}\Big)^q+
3\Big(\frac{1+\lambda}{6}\Big)^q\Big]\,,\\
{\cal I}_q(A:B) &=& \frac{1}{q-1}\Big[\frac12(1-\lambda)^q+\frac32\Big(\frac{1+\lambda}{3}\Big)^q-2^{1-q}\Big]
\,.
\end{eqnarray}
Finally, both $q$-discords for Werner states turn out to be 
\begin{eqnarray}
\widetilde{D}_q(A:B) \!\!&=&\!\! \frac{1}{q-1}\Big[1-2^{1-q}+\Big(\frac{1-\lambda}{2}\Big)^q
+3\Big(\frac{1+\lambda}{6}\Big)^q-\Big(\frac{2-\lambda}{3}\Big)^q-\Big(\frac{1+\lambda}{3}\Big)^q\Big]\,,\\
{D}_q(A:B) \!\!&=&\!\! \frac{1}{q-1}\Big[\frac12(1-\lambda)^q+
\frac{1}{2}\Big(\frac{1+\lambda}{3}\Big)^q-\Big(\frac{2-\lambda}{3}\Big)^q\Big]\,.
\end{eqnarray}
\begin{figure}[t]
\centerline{\includegraphics[height=6cm]{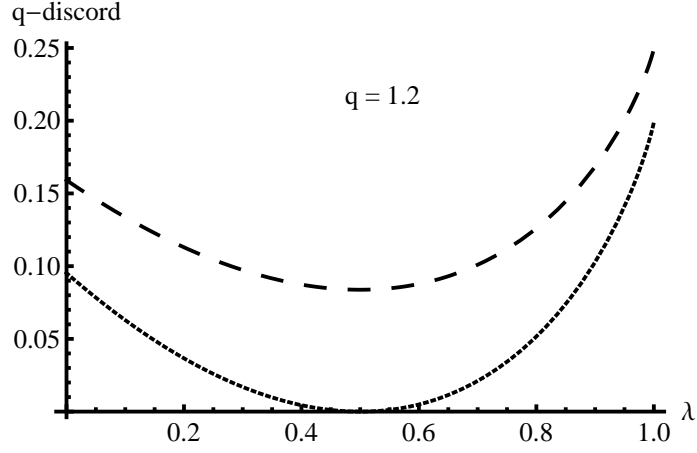}}
\caption{\label{fid-dwa} Comparison between two kinds of $q$-discord
for Werner states for $q=1.2$. The upper graph corresponds to $\widetilde{D}_q(A:B)$ whereas 
the lower to $D_q(A:B)$.}
\end{figure}
The comparison between $D_q(A:B)$ and $\widetilde{D}_q(A:B)$ for $q=1.2$ is shown in 
Fig.~\ref{fid-dwa}. Note, that for classically correlated (diagonal) Werner state corresponding to $\lambda=1/2$, 
$D_q(A:B)=0$ for every $q>0$,  as one should expect.
The quantity $\widetilde{D}_{1.2}(A:B)$ is positive for this state and for $0<q<1$ 
can be even negative which makes it less useful 
as a measure of quantum correlations. 

We can prove (see Appendix) that the $q$-discord $D_q(A:B)$ for $2\otimes 2$ Werner states is always nonnegative for $q>0$.

Similar calculations can be carried out also for $2\otimes 2$ isotropic states defined as 
\[
\rho_{\rm iso}(\lambda)=\lambda \Pi_++\frac{1-\lambda}{3}(I_4-\Pi_+)\,,\qquad 0\leq\lambda\leq 1,
\]
where $\ket{\psi_+}=1/\sqrt{2}(\ket{00}+\ket{11})$ and the projector $\Pi_+=\ket{\psi_+}\bra{\psi_+}$.
This family is invariant with respect to the following action of local unitary transformations:
\[ U\otimes U^*\rho_{\rm iso}(\lambda)\, (U\otimes U^*)^\dagger=\rho_{\rm iso}(\lambda)\,.
\]
After straightforward algebra one obtains two types of $q$-discords for isotropic states 
as
\begin{eqnarray}
\widetilde{D}_q(A:B) \!\!&=&\!\! \frac{1}{q-1}\Big[1-2^{1-q}+3\Big(\frac{1-\lambda}{3}\Big)^q
+\lambda^q-\Big(\frac{2-2\lambda}{3}\Big)^q-\Big(\frac{1+2\lambda}{3}\Big)^q\Big]\,,\\
{D}_q(A:B) \!\!&=&\!\! \frac{1}{q-1}\Big[\frac12\Big(\frac{2-2\lambda}{3}\Big)^q+
\frac{1}{2}(2\lambda)^q-\Big(\frac{1+2\lambda}{3}\Big)^q\Big]\,.
\end{eqnarray}
\begin{figure}[t]
\centerline{\includegraphics[height=6cm]{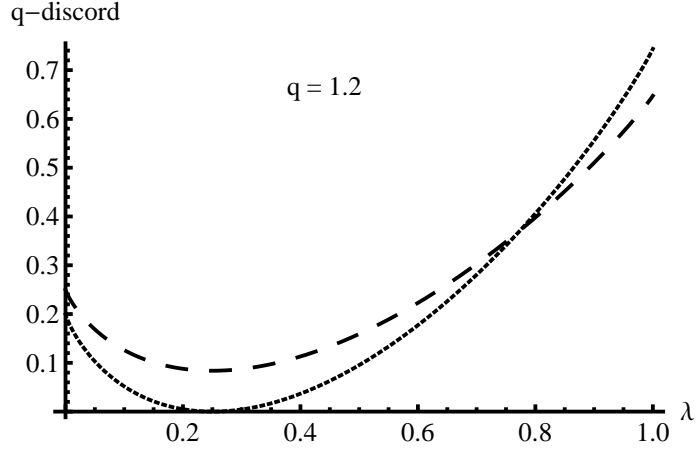}}
\caption{\label{fig-dwa-iso} Comparison between two kinds of $q$-discord
for isotropic states for $q=1.2$. The dashed-line graph corresponds to $\widetilde{D}_q(A:B)$ whereas 
the dotted-line to $D_q(A:B)$.}
\end{figure}
The comparison between $D_q(A:B)$ and $\widetilde{D}_q(A:B)$ for $q=1.2$ is shown in 
Fig.~\ref{fig-dwa-iso}. Now, the classically correlated isotropic state corresponds to $\lambda=1/4$ and 
as expected, for this value of $\lambda$, $D_q(A:B)=0$ for every $q>0$. 
We can prove that the $q$-discord $D_q(A:B)$ for $2\otimes 2$ isotropic states is always nonnegative 
for every $q>0$.

The quantity $\widetilde{D}_{1.2}(A:B)$ is positive for this state and for $0<q<1$ 
can be even negative which makes its behaviour similar to that of Werner states.

\section{Circulant Two-Qubit States}

As a final example, let us consider the following family of circulant two-qubit states:
\begin{equation}\label{e,g-states}
\rho_{\varepsilon,g}\;=\;\frac{1}{2+\varepsilon+\varepsilon^{-1}}\left[\begin{array}{cccc}
1 & 0 & 0 & 1 \\
0 &\varepsilon & g & 0 \\
0 & g & \varepsilon^{-1} & 0 \\
1 & 0 & 0 & 1
\end{array}\right],\qquad 0\leq\varepsilon,g\leq 1\,.
\end{equation}
The family does not belong  neither to the Bell diagonal states considered in \cite{Luo1} nor
to its generalisation investigated in \cite{Fei}. But the method of discord calculation 
used in \cite{Ali,Fei} can be adopted for $q$-discord $D_q(A:B)$ as well.

One can immediately check that eigenvalues of $\rho_{\varepsilon,g}$ read
\[
\lambda_0=0\,,\quad \lambda_1=\frac{2\varepsilon}{(1+\varepsilon)^2}\,,\quad
\lambda_{3,4}=\frac{1+\varepsilon^2\pm\sqrt{\varepsilon^4+2(2g^2-1)\varepsilon^2+1}}{2(1+\varepsilon)^2}\,,
\] 
whereas eigenvalues of reductions $\rho_A$ and $\rho_B$ are the following:
\[ 
\mu_1\;=\;\frac{1+\varepsilon}{2+\varepsilon+\varepsilon^{-1}}\,,\qquad
\mu_2\;=\;\frac{1+\varepsilon^{-1}}{2+\varepsilon+\varepsilon^{-1}}\,.
\]
Therefore,
\[
T_q(A,B)\;=\;\frac{1}{q-1}\Big[1-\sum_{k=1}^4\lambda_k^q\Big]
\]
and
\[
T_q(A)\;=\;T_q(B)\;=\;\frac{1}{q-1}[1-\mu_1^q-\mu_2^q]\,.
\]
Then, the quantum mutual Tsallis information (\ref{mut-inf-3b}) results in
\begin{eqnarray*}
&&\hspace*{-10mm}{\cal I}_q(A:B)\;=\;\\
&&=\;\frac{\varepsilon^q(2^q-\varepsilon^q-2)-1+[\frac12(1+\varepsilon^2)+
\frac12\sqrt{\varepsilon^4+2(2g^2-1)\varepsilon^2+1}]^q}{(q-1)(1+\varepsilon^q)(1+\varepsilon)^q}\\
&&\hspace*{20mm}
+\;\frac{[\frac12(1+\varepsilon^2)-
\frac12\sqrt{\varepsilon^4+2(2g^2-1)\varepsilon^2+1}]^q}{(q-1)(1+\varepsilon^q)(1+\varepsilon)^q}\,.
\end{eqnarray*}

Now, in order to obtain classical correlations ${\cal C}_q(A:B)$ we parameterise the 
projective measurement $\Pi=\{\Pi_1,\Pi_2\}$ by $2\times 2$ unitary matrix $U$
\[ \Pi_k\;=\;U\ket{k}\bra{k}U^*\,,\quad k=0,1\,,\quad U\in{\rm U}(2)
\]
of the form 
\[ U\;=\;t_0\1+i\vec{t}\cdot\vec{\sigma}\;=\;
\left[\begin{array}{cc}
t_0+t_3 & t_1-it_2 \\
t_1+it_2 & t_0-t_3
\end{array}\right]
\,,\quad t_0^2+t_1^2+t_2^2+t_3^2=1\,,
\]
where $\vec{\sigma}=(\sigma_1,\sigma_2,\sigma_3)$ are Pauli matrices. The ensemble of post-measurement states
$\{\rho_k,p_k\}$, $k=0,1$, can be characterized
by probabilities (see \cite{Ali} for details) 
\begin{eqnarray}
p_0 &=& (1+\varepsilon^{-1})k+(1+\varepsilon)l\,,\\
p_1 &=& (1+\varepsilon^{-1})l+(1+\varepsilon)k\,,
\end{eqnarray}
and the eigenvalues of $\rho_0$ and $\rho_1$ (see (\ref{rho_k}))
\begin{equation}\label{spektrum}
{\rm Spec}(\rho_0)=\Big\{\frac{1}{2}(1\pm\vartheta)\Big\}\,,\qquad
{\rm Spec}(\rho_1)=\Big\{\frac{1}{2}(1\pm\vartheta')\Big\}\,,
\end{equation}
where 
\begin{eqnarray*}
\vartheta &=& \sqrt{\frac{[(1-\varepsilon^{-1})k+(\varepsilon-1)l]^2+4kl(1+g)^2-16mg}{
[(1+\varepsilon^{-1})k+(\varepsilon+1)l]^2}}\,, \\
\vartheta' &=& \sqrt{\frac{[(1-\varepsilon^{-1})l+(\varepsilon-1)k]^2+4kl(1+g)^2-16mg}{
[(1+\varepsilon^{-1})l+(\varepsilon+1)k]^2}}\,,
\end{eqnarray*}
with $k\in[0,1]$, $l=1-k$, $m\in [0,1/4]$ defined in terms of $(t_0,\vec{t})$ as
\[
k=t_0^2+t_3^2\,,\qquad l=t_1^2+t_2^2\,,\qquad m=(t_0t_1+t_2t_3)^2\,.
\] 
The quantum conditional Tsallis entropy with respect to the measurement yields
\begin{equation}\label{cond-T-entropy}
T_q(\rho_{\varepsilon,g}|\{\Pi_k\})\;=\;
\frac{p_0^qT_{q}(\rho_0)+p_1^qT_q(\rho_1)}{p_0^q+p_1^q}
\end{equation}
and 
\begin{equation}\label{class-1}
{\cal C}_q(A:B)\;=\;T_q(A)-\inf_{\Pi_k}T_q(\rho_{\varepsilon,g}|\{\Pi_k\})\,.
\end{equation}
Let us define the function 
\[
f_q(x)\;=\;\Big[1-\Big(\frac{1+x}{2}\Big)^q-\Big(\frac{1-x}{2}\Big)^q
\Big]\,.
\]
which is decreasing in $x$ for $q>0$.  
Due to \cite{Ali} the infimum in (\ref{class-1}) can be obtained when:
\begin{itemize}
\item[1)] $k=0$ and $l=1$. This results in $m=0$, $p_0=\varepsilon+1$, 
$p_1=1+\varepsilon^{-1}$ and 
\[
\vartheta=\vartheta'=\frac{\varepsilon-1}{\varepsilon+1}\,.
\]
From (\ref{spektrum}) and (\ref{cond-T-entropy}) one obtains
\begin{eqnarray}
 T_q(\rho_{\varepsilon,g}|\{\Pi_k\}) &=& T_{q}(\rho_0)\;=\;T_q(\rho_1)\nonumber \\
 &=& f_q(\vartheta)\;=\;\frac{1}{q-1}\Big[1-\frac{\varepsilon^q+1}{(\varepsilon+1)^q}\Big]\,.\label{eq1}
\end{eqnarray}
\item[2)] $k=1$ and $l=0$. This results in the same formula (\ref{eq1})
\item [3)] $k=l=\frac12$. This results in $m=0$ or $m=1/4$, hence
\[
 \vartheta\;=\;\vartheta'\;=\;\left\{\begin{array}{cl}
\dST\frac{\sqrt{(\varepsilon-\varepsilon^{-1})^2+4(1+g)^2}}{2+\varepsilon+\varepsilon^{-1}} &\quad 
m=0 \\[2ex]
\dST\frac{\sqrt{(\varepsilon-\varepsilon^{-1})^2+4(1-g)^2}}{2+\varepsilon+\varepsilon^{-1}} &\quad \dST
m=\frac{1}{4}\,,
\end{array}\right.
\]
and 
\begin{eqnarray}
 T_q(\rho_{\varepsilon,g}|\{\Pi_k\}) &=& T_{q}(\rho_0)\;=\;T_q(\rho_1)\;=\;f_q(\vartheta)\,.
\end{eqnarray}
\end{itemize}
Note that
\[
\frac{\varepsilon-1}{\varepsilon+1}=\frac{\varepsilon-\varepsilon^{-1}}{2+\varepsilon
+\varepsilon^{-1}}\leq\frac{\sqrt{(\varepsilon-\varepsilon^{-1})^2+4(1-g)^2}}{2+\varepsilon+\varepsilon^{-1}}
\leq\frac{\sqrt{(\varepsilon-\varepsilon^{-1})^2+4(1+g)^2}}{2+\varepsilon+\varepsilon^{-1}}\,,
\]
therefore
\[
\inf_{\Pi_k}T_q(\rho_{\varepsilon,g}|\{\Pi_k\})\;=\;f_q\Big(\frac{\sqrt{(\varepsilon-\varepsilon^{-1})^2+4(1+g)^2}}{2+\varepsilon+\varepsilon^{-1}}\Big)\
\]
and (\ref{class-1}) takes the form 
\begin{equation}\label{class-2}
{\cal C}_q(A:B)\;=\;\frac{1}{q-1}\Big[1-\Big(\frac{1+\varepsilon}{2+\varepsilon+\varepsilon^{-1}}\Big)^q-\Big(\frac{1+\varepsilon^{-1}}{2+\varepsilon+\varepsilon^{-1}}\Big)^q\Big]-f_q\Big(\frac{\sqrt{(\varepsilon-\varepsilon^{-1})^2+4(1+g)^2}}{2+\varepsilon+\varepsilon^{-1}}\Big).
\end{equation}

\begin{figure}[t]
\centerline{\includegraphics[height=6cm]{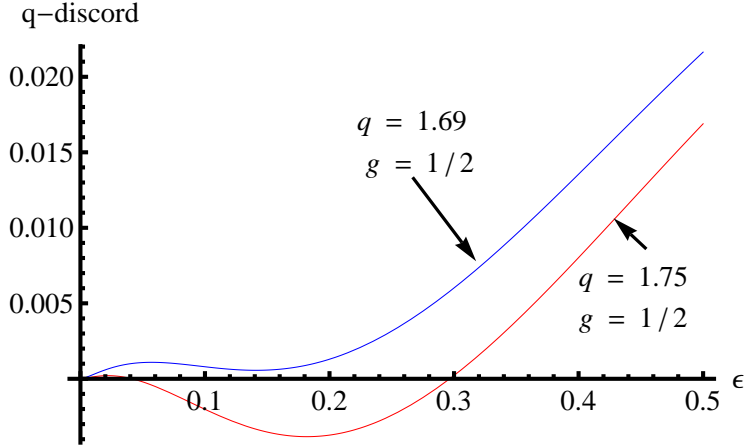}}
\caption{\label{ujemny} $q$-Discord for $(g,\varepsilon)$-states. For $q=1.75$ and $g=0.5$ $q$-discord 
$D_q(A:B)$ is negative for some interval of $\varepsilon$.}
\end{figure}

As noticed in \cite{Coles,Ros}, the $q$-discord  $\widetilde{D}_q(A:B)$ can take negative values for some states and 
$q>2$. Now, we discover the same for $D_q(A:B)$ for some specific $(g,\varepsilon)$-states and $q=1.75$ (see Fig.~\ref{ujemny}).

\section*{Acknowledgments}

The author would like to acknowledge many stimulated discussions with Prof. D. Chru\'sci\'nski.

\section*{Appendix}

We have the following simple lemma
\begin{lemma}
If  $x+y=n$ for $x,y\geq 0$, then 
\begin{itemize}
\item $\displaystyle x^q+y^q\geq 2\Big(\frac{n}{2}\Big)^q$ for $q\geq 1$,
\item $\displaystyle x^q+y^q\leq 2\Big(\frac{n}{2}\Big)^q$ for $0<q<1$
\end{itemize}
\end{lemma}
The proof results straightforward from the obvious fact that the function $f(x)=x^q+(n-x)^q$ takes 
its local minimum (maximum) at $x=n/2$ for $q>1$ ($0<q<1$). At this point $f(n/2)=2(n/2)^q$.

Note,  that for Werner states the inequality $D_q(A:B)\geq 0$ is equivalent to
\[ \Big(\frac{3(1-\lambda)}{2-\lambda}\Big)^q+\Big(\frac{1+\lambda}{2-\lambda}\Big)^q\geq 2
\quad\mbox{for}\quad q>1
\]
or
\[ \Big(\frac{3(1-\lambda)}{2-\lambda}\Big)^q+\Big(\frac{1+\lambda}{2-\lambda}\Big)^q\leq 2
\quad\mbox{for}\quad 0<q<1\,.
\]
Evidently
\[ 
\Big(\frac{3(1-\lambda)}{2-\lambda}\Big)+\Big(\frac{1+\lambda}{2-\lambda}\Big)=2\,,
\]
hence we can apply Lemma~1 which proves nonnegativity of $D_q(A:B)$ for Werner states.
The same method can be used to show that also for isotropic states the same conclusion holds.

\end{document}